\documentclass[11pt]{article}
\usepackage{pdfsync}
\usepackage{graphicx} 
\usepackage{subfig}
\usepackage{color}
\usepackage{amsfonts}
\usepackage{amssymb}
\usepackage{epsfig}
\usepackage{dcolumn}
\usepackage{booktabs}
\usepackage{multirow}
\usepackage{makecell}
\usepackage{boldline}
\usepackage{marvosym}

\usepackage[colorlinks]{hyperref}
\hypersetup{
    colorlinks=true,
    linkcolor=blue,
    filecolor=magenta,  
    urlcolor=[rgb]{0.00,0.00,0.50},    
    citecolor=red,
    }

\usepackage{url}

\advance \textheight by \topskip
\usepackage{amsmath}
\usepackage[english]{babel}
\makeatletter
 \@addtoreset{equation}{section}
\makeatother

\usepackage{amsmath,amssymb}
\makeatletter \@addtoreset{equation}{section}
\makeatother
\def\be{\begin{equation}}

\def\be{\begin{equation}}
\def\ee{\end{equation}}

\def\Z{\mathbb Z}
\def\R{\mathbb R}

\def\C{\mathbb C}

\usepackage{amsmath,amssymb}
\def\bea{\begin{eqnarray}}
\def\eea{\end{eqnarray}}
\def\barray{\begin{array}}
\def\earray{\end{array}}

\begin{document}

\title{
{\bf 
Ground-state isolation and discrete  flows 
 in a rationally extended quantum harmonic oscillator\\
}}

\author{{\bf Jos\'e F. Cari\~nena${}^a$ and Mikhail S. Plyushchay${}^b$}  \\
[8pt]
{\small \textit{
${}^a$Departamento de F\'{\i}sica Te\'orica, 
Universidad de Zaragoza, 50009 Zaragoza, Spain}}\\
{\small \textit{ ${}^b$Departamento de F\'{\i}sica,
Universidad de Santiago de Chile, Casilla 307, Santiago 2,
Chile  }}\\
[4pt]
 \sl{\small{E-mails:  \textcolor{blue}{jfc@unizar.es}, 
\textcolor{blue}{mikhail.plyushchay@usach.cl}
}}
}
\date{}
\maketitle

\begin{abstract}

Ladder operators for the simplest version of 
a rationally extended  quantum harmonic oscillator
(REQHO) 
are constructed by  applying a Darboux transformation 
to the quantum harmonic oscillator system.
It is shown that the physical spectrum of the REQHO carries 
a direct sum of a trivial  and an infinite-dimensional 
 irreducible representation 
of the polynomially  deformed bosonized 
 $\mathfrak{osp}(1|2)$ superalgebra.
In correspondence with this the ground state of the system 
is isolated from other physical states but 
can be reached  by  ladder operators 
 via non-physical energy eigenstates, 
which belong to either an infinite chain
of similar eigenstates
or to the chains with  generalized Jordan states. 
We show that the discrete  
chains of the states
generated by 
 ladder operators and 
associated with physical energy levels 
include six basic generalized Jordan states,
in comparison with the two basic Jordan states entering
in 
analogous discrete chains for 
the quantum harmonic oscillator.
\end{abstract}

\vskip.5cm\noindent

\section{Introduction}
Darboux transformations, 
introduced originally as a method to solve 
linear differential equations and generalized 
subsequently for Darboux-Crum(-Krein-Adler)
 transformations \cite{Darb,Crum,Krein,Adler,Matveev}, 
find many important applications in physics.  
For a long period of time they were  
used in quantum mechanics 
in  factorization method 
for solving Schr\"odinger equation \cite{Schr,InfHull,CR00,MelRos}.
Nowadays they are exploited intensively 
in the context of supersymmetry.
These transformations lie in the heart of
supersymmetric quantum 
mechanics \cite{Witten1,Witten2,CoKhSu,CR01,CR08}.
They  are particularly  employed 
for the construction of  new solvable 
and quasi-exactly solvable quantum 
mechanical systems.
The Darboux transformations   play a fundamental role in investigation
of nonlinear equations in partial derivatives 
and partial difference equations, 
where  they allow to relate different integrable 
systems and provide an effective method 
for the construction of  nontrivial solutions for them
\cite{Matveev,ArPl}.
Periodic Darboux chains generate 
finite-gap systems \cite{VesSha}
in an  alternative  way to the original 
algebro-geometric approach \cite{NMPZ,GesHol}. 
Such  chains generate also the quantum
harmonic oscillator (QHO) system and  
Painlev\'e equations \cite{VesSha,AdlPai,Hiet}, 
which are  intimately related with isomonodromic 
deformations of linear systems and integrability properties
of nonlinear systems in partial derivatives.
Recently, the isomonodromic deformations \cite{Iso1,Iso2}
and Darboux transformations  played a 
key role in the discovery and investigation
of the properties of the new class  of 
exceptional orthogonal polynomials
 \cite{Adler,Dub,CPRS,FellSmi,Exc1,OdSas,SasTsZh,Ques,Grand,Sesma,GGM,Pupas}.  
One of such a family corresponds 
to exceptional Hermite polynomials, 
which can be obtained by applying Darboux and Darboux-Crum
transformations to the QHO system. 
The quantum mechanical systems appearing in such a way
are described by certain rational extensions of the harmonic potential.

A  simplest rationally extended quantum harmonic oscillator 
(REQHO) \cite{Dub,CPRS}
can be obtained from the QHO system by
applying to it Darboux transformation generated by 
the  ``Wick-rotated" second excitation of the ground-state.
The resulting system is characterized 
by an infinite tower of equidistant bound states 
which are separated from the  ground-state 
by a triple energy gap. As a consequence, 
the general solution of 
the evolution problem for REQHO,
like for the quantum harmonic  and isotonic oscillators,  
is periodic in time with 
a constant (not depending on energy) period
\cite{Dub,AsoCar,CPR07}.
For the discussion of different 
aspects of this quantum mechanical system see Refs.
\cite{Adler,Dub,CPRS,FellSmi,Sesma,GGM,Pupas}.

It is known that 
the spectrum of the QHO carries an infinite-dimensional irreducible 
representation of the bosonized $\mathfrak{osp}(1\vert 2)$ 
superalgebra which can be  generated by means of  
the  creation and annihilation operators
identified as fermionic generators 
\cite{CromRit,defOSP}~\footnote{For some recent investigations
on  superconformal quantum mechanical symmetry and its applications see
\cite{KriLech,FedLuk,PlyWi,TDB,IST}. }.
In this context there appears a rather natural question: what are the
ladder operators in rationally extended 
quantum harmonic oscillator systems and 
what spectrum  generating 
algebras do they produce? 
In this paper we answer these questions 
for the simplest case of the REQHO system
by employing the properties of the 
Darboux transformations. 
A special 
role in the construction we obtain 
 belongs to
 generalized Jordan states. 

The  
paper is organized as follows. 
In the next section we review general properties 
of the Darboux transformations 
 and 
related Jordan states.
In Section 3 we discuss discrete flows generated by
the ladder operators in  the QHO system 
and recall the bosonized superconformal $\mathfrak{osp}(1\vert 2)$ 
structure appearing in it in the form
of the spectrum generating superalgebra.
In Section 4 we generate the simplest REQHO  
by applying Darboux transformation to the QHO.
Then we construct ladder operators  for REQHO
by a Darboux-dressing of the creation and annihilation operators
of the QHO system, 
consider discrete flows  and 
discuss a polynomially deformed   bosonized 
$\mathfrak{osp}(1\vert 2)$ structure
in the REQHO that reflects in a coherent way
peculiarities of its spectrum.
Section 5 is devoted to the conclusion and outlook. 
 In Appendix we describe the action 
of the ladder operators  on non-physical eigenstates 
of the quantum harmonic oscillator which are closely related to its
physical spectrum,  and  present the construction 
of the net of associated Jordan and generalized Jordan states. 

\section{Darboux transformations 
and Jordan states}

Let $\psi_*(x)$ be a 
solution of the stationary Schr\"odinger equation 
$H\psi_*=E_*\psi_*$.  For the moment  we consider this equation 
formally as an abstract second order differential equation 
in which   $H=-\frac{d^2}{dx^2}+V(x)$ is a  Hamiltonian operator 
with a real nonsingular on $\R$ potential $V(x)$. A real constant  $E_*$
is treated  here as an eigenvalue
without preoccupying the questions of boundary conditions 
and normalizability for $\psi_*(x)$.
Consequently, we 
do not distinguish functions  $\psi(x)$ and 
$C\psi(x)$,  where $C\in\C$, $C\neq 0$,
and  assume  that modulo such a multiplicative factor
$\psi(x)$  is  chosen to be a  real-valued  function.
A linearly independent solution 
for the same eigenvalue $E_*$ can be taken in the form 
\be\label{secsol}
\widetilde{\psi_*}(x)=\psi_*(x)\int^x\frac{d\xi}{\psi^2_*(\xi)}\,.
\ee
Due to integration with an indefinite lower limit,
function  $\widetilde{\psi_*}(x)$ 
is supposed to be defined up to  an
additive term proportional to ${\psi_*}(x)$.
Assume  now that $E_*$ is chosen so that 
function $\psi_*(x)$ is 
nodeless, $\psi_*(x)\neq 0$,
and introduce the first-order 
differential operators 
\begin{equation}\label{Adef}
A_{\psi_*}=\psi_*\frac{d}{dx}\frac{1}{\psi_*}=
\frac{d}{dx}-\mathcal{W}(x)\,,\qquad 
\mathcal{W}(x)= \frac{\psi'_*}{\psi_*} \,,
\end{equation}
and $A_{\psi_*}^\dagger=-\frac{d}{dx}-\mathcal{W}(x)$,
where  prime denotes derivative in $x$.
Note that as $A_{\psi_*}$ and $A_{\psi_*}^\dagger$ are
first-order differential operators, their kernels are one-dimensional,
\begin{equation}
	\ker\, A_{\psi_*}= {\psi_*}, 
	\qquad \ker\, A^\dag_{\psi_*}=\frac 1{\psi_*}\,.
	\label{kernels}
\end{equation} 
These operators
provide a factorization of the shifted for the
 constant $E_*$  Hamiltonian,
$H-E_*=A_{\psi_*}^\dagger A_{\psi_*}$.
Potential $V(x)$   and  superpotential $\mathcal{W}(x)$
are connected by a relation 
$V(x)-E_*=\mathcal{W}^2+\mathcal{W}'$. 
The product with the permuted  first-order operators,
$A_{\psi_*} A_{\psi_*}^\dagger=\breve{H}-E_*$,
defines the associated partner  system described by the Hamiltonian 
$\breve{H}=-\frac{d^2}{dx^2}+\breve{V}(x)$ with 
$\breve{V}(x)-E_*=\mathcal{W}^2-\mathcal{W}'$. 
{}From the alternative factorization relations it follows immediately   that 
the first-order operators $A_{\psi_*}$ and $A_{\psi_*}^\dagger $ 
intertwine quantum  Hamiltonians 
$H$ and $\breve{H}$,
\be\label{intertrel}
A_{\psi_*} H=\breve{H}A_{\psi_*}\,,\qquad
A_{\psi_*}^\dagger\breve{H}=HA_{\psi_*}^\dagger\,.
\ee
If $\psi_{{}_{E}}$ is a physical (normalizable)  or non-physical (non-normalizable)
 solution of the Schr\"odinger equation
$H\psi_{{}_{E}}=E\psi_{{}_{E}}$ for some 
eigenvalue $E\neq E_*$, then as a consequence of (\ref{intertrel}), 
$A_{\psi_*}\psi_{{}_{E}}$ will be 
an eigenstate of
$\breve{H}$ of the same,  physical or non-physical,  nature 
for the same eigenvalue $E$\,:
$\breve{H}(A_{\psi_*}\psi_{{}_{E}})=E(A_{\psi_*}\psi_{{}_{E}})$.
Particularly, for the linear independent solution
$\widetilde{\psi_{{}_{E}}}$  constructed from $\psi_{{}_{E}}$
according to the rule (\ref{secsol}),
$\widetilde{\psi_{{}_{E}}}=\psi_{{}_{E}}(x)\int^x d\xi/\psi^2_{{}_{E}}(\xi)$,
we have $\breve{H}(A_{\psi_*}\widetilde{\psi_{{}_{E}}})=E(A_{\psi_*}\widetilde{\psi_{{}_{E}}})$.
On the other hand, for $E=E_*$ and $\psi_{{}_{E}}=\widetilde{\psi_*}$ we find that
\be\label{Apsi*}
 A_{\psi_*}\widetilde{\psi_*}=\frac{1}{\psi_*}\,. 
 \ee
The function $\frac{1}{\psi_*}$
 is  the  kernel of the operator $A_{\psi_*}^\dagger$,
and  therefore is an eigenstate
of 
$\breve{H}$, $(\breve{H}-E_*)\frac{1}{\psi_*}=0$. 
Analogously, if $\breve{\psi}_{{}_{E}}$ 
is an eigenfunction  of $\breve{H}$ of eigenvalue $E\neq E_*$, then 
$A_{\psi_*}^\dagger\breve{\psi}_{{}_{E}}$ is an eigenstate 
of $H$ of the same eigenvalue,
$H(A_{\psi_*}^\dagger\breve{\psi}_{{}_{E}})=E(A_{\psi_*}^\dagger \breve{\psi}_{{}_{E}})$.
For $E=E_*$ the application of $A_{\psi_*}^\dagger$ to 
a  linearly independent from ${\frac{1}{\psi_*}}$
eigenfunction  $\widetilde{\left(\frac{1}{\psi_*}\right)}$ of $\breve{H}$
maps it into  the kernel of $A_{\psi_*}$,
\be\label{A+psi*}
A_{\psi_*}^\dagger\widetilde{\left(\frac{1}{\psi_*}\right)}=\psi_*\,,
\ee
that is the eigenstate of $H$.

\begin{figure}[htbp]
 \begin{center}
\includegraphics[scale=0.3]{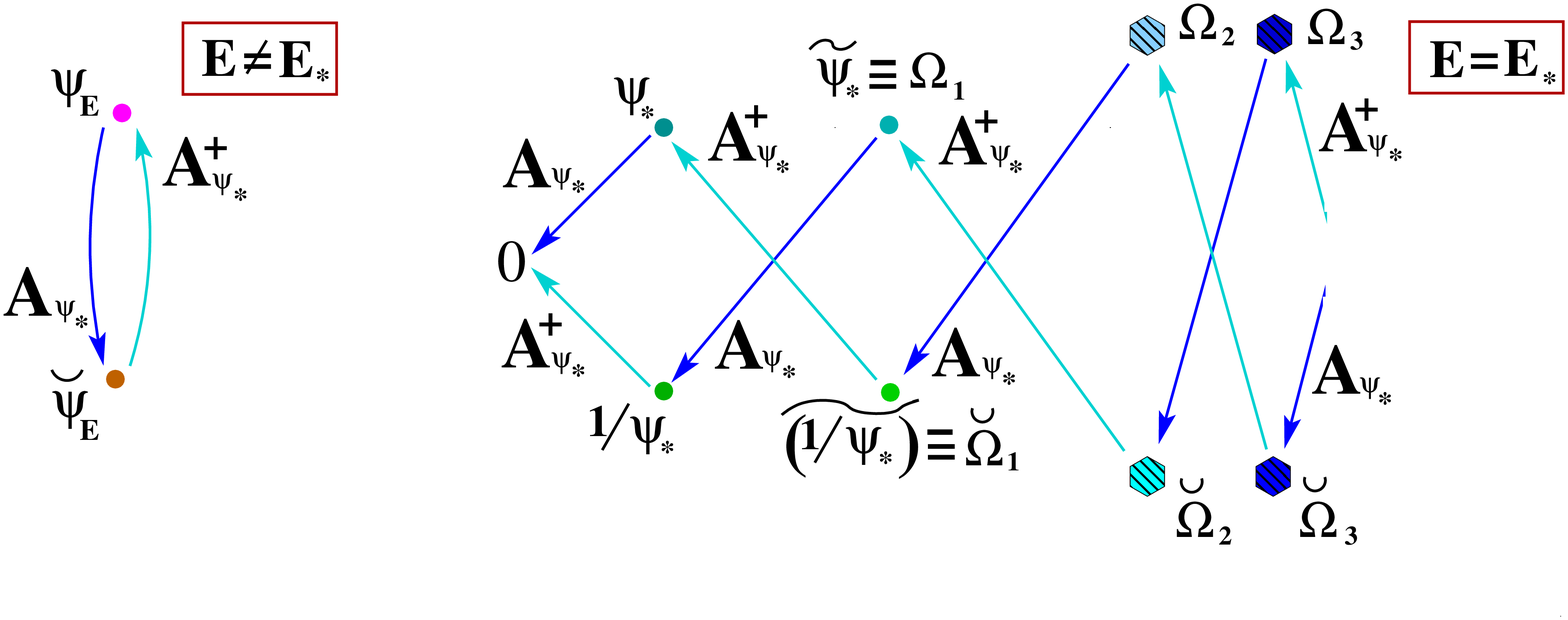}
\caption{ 
Action of the Darboux transformation generators. }
\label{Fig1}
\end{center}
\end{figure}

The described structure of the 
Darboux transformations reveals an essential difference between the 
cases  $E\neq E_*$ and  $E= E_*$.
 The action of the Darboux transformation 
generators on the eigenstates with  $E\neq E_*$ is of the two-cyclic
nature  in the 
following sense: 
if  $\psi$ is such that $H\psi=E\psi$, then 
$A_{\psi_*}$ maps this state into an eigenstate of $\breve{H}$, 
$A_{\psi_*}\psi=\breve{\psi}$, $\breve{H}\breve{\psi}=E\breve{\psi}$,
while application of $A_{\psi_*}^\dagger$ to $\breve{\psi}$ 
reproduces (up to a multiplicative factor)  the initial state $\psi$.  
At the same time, for $E=E_*$ we have  
$A_{\psi_*}\widetilde{\psi_*}=\frac{1}{\psi_*}$, 
$A_{\psi_*}^\dagger \frac{1}{\psi_*}=0$ and 
$A_{\psi_*}^\dagger \widetilde{ \left(\frac{1}{\psi_*}\right)}=\psi_*$, 
$A_{\psi_*}\psi_*=0$, and no analogous cyclic structure
does appear.
In this case the functions 
\be\label{Omega}
\Omega_2(x)={\psi_*(x)}\int^x\frac{1}{\psi_*(\xi)}\widetilde{ \left(\frac{1}{\psi_*}\right)}(\xi)d\xi\,,
\qquad
\breve{\Omega}_2(x)=\frac{1}{\psi_*(x)}\int^x\psi_*(\xi)\widetilde{\psi_*}(\xi)d\xi\,
\ee
are the pre-images  of  $ \widetilde{ \left(\frac{1}{\psi_*}\right)}$ 
and $\widetilde{\psi_*}$,
\be\label{AA+}
A_{\psi_*}\Omega_2=\widetilde{ \left(\frac{1}{\psi_*}\right)}\equiv\breve{\Omega}_1\,,
\qquad
A_{\psi_*}^\dagger \breve{\Omega}_2=\widetilde{\psi_*}\equiv \Omega_1\,.
\ee
Similarly to  the eigenstates   of the form (\ref{secsol}), 
functions $\Omega_2(x)$ and $\breve{\Omega}_2(x)$ are 
defined  modulo additive terms ${\psi_*(x)}$ and 
$\frac{1}{\psi_*(x)}$, respectively. 
Wave functions  (\ref{Omega}) are not, however, 
formal  eigenfunctions of 
$H$ and $\breve{H}$, but as a consequence of 
(\ref{AA+}) they obey the relations 
$A_{\psi_*} A_{\psi_*}^\dagger A_{\psi_*}\Omega_2=0$
and
$A_{\psi_*}^\dagger A_{\psi_*}A_{\psi_*}^\dagger\breve{\Omega}_2=0$.
Therefore,
\be\label{Jordan}
\qquad({H}-E_*)^2\,\Omega_2=0\,,\qquad
(\breve{H}-E_*)^2\, \breve{\Omega}_2=0\,,
\ee
and we conclude that 
 $\Omega_2$ and $\breve{\Omega}_2$ 
are  generalized  eigenstates
of $H$ and $\breve{H}$  of rank $2$ corresponding to the
same eigenvalue $E=E_*$.
Having in mind a  generalization
of relations of the form (\ref{Jordan}) which appear in the following,
see particularly 
 Eq. (\ref{chi-Jordan}) below, 
we  refer to $\Omega_2$ and $\breve{\Omega}_2$  
as Jordan states of order $2$  \cite{Jord1,Jord2}.
The states (\ref{Omega}) 
can be generalized further  by defining 
\be\label{Omegan}
\Omega_n(x)={\psi_*(x)}\int^x\frac{1}{\psi_*(\xi)}\breve{\Omega}_{n-1}(\xi)d\xi\,,
\qquad
\breve{\Omega}_n(x)=\frac{1}{\psi_*(x)}\int^x\psi_*(\xi)\Omega_{n-1}(\xi)d\xi\,,
\ee
where $n=2,3,\ldots$.
These states obey the relations 
$A_{\psi_*}\Omega_n=\breve{\Omega}_{n-1}$,
$A_{\psi_*}^\dagger\breve{\Omega}_n=\Omega_{n-1}$.
Consequently we find that $\Omega_n$ and 
$\breve{\Omega}_n$ are annihilated by 
differential operators of order $n+1$ constructed in terms of 
$A_{\psi_*}$ and $A_{\psi_*}^\dagger$.  Namely, 
for even $n=2k$  we have $A_{\psi_*}(A_{\psi_*}^\dagger A_{\psi_*})^k\Omega_{2k}=0$, 
$A_{\psi_*}^\dagger(A_{\psi_*} A_{\psi_*}^\dagger)^k\breve{\Omega}_{2k}=0$,
while for odd $n=2k-1$ we obtain
$(A_{\psi_*}^\dagger A_{\psi_*})^k\Omega_{2k-1}=0$
and $(A_{\psi_*} A_{\psi_*}^\dagger)^k\breve{\Omega}_{2k-1}=0$, $k=1,\ldots$.
In both cases of the even and odd 
values of $n$ we have
$(H-E_*)^{n+1}\Omega_{2n}=0$, 
$(\breve{H}-E_*)^{n+1}\breve{\Omega}_{2n}=0$,
$n=1,2,\ldots$,
and $(H-E_*)^{n+1}\Omega_{2n+1}=0$,
$(\breve{H}-E_*)^{n+1}\breve{\Omega}_{2n+1}=0$,
$n=0,1,\ldots$.  Thus, $\Omega_k$ and 
$\breve{\Omega}_k$ with $k=2n,\,  2n+1$ are
Jordan states of order $n+1$. 
The described properties 
are illustrated by Figure \ref{Fig1}.

 \vskip0.3cm

\section{Discrete flows in harmonic oscillator system
and the $\mathfrak{osp}(1\vert 2)$ }

Let us consider  now the QHO system with $V(x)=x^2$.
Its physical bound eigenstates with eigenvalues $E_n=2n+1$ 
are described by normalizable wave functions
$\psi_n(x)=H_n(x)e^{-x^2/2}$, $n=0,1,\ldots$, where $H_n(x)$ 
are the Hermite polynomials.
The change  $x\rightarrow ix$ transforms Hamiltonian  of the QHO
\begin{equation}\label{HamQHO}
H=-\frac{d^2}{dx^2}+x^2
\end{equation}
 into $-H$, from where it follows that 
the wave functions $\psi^-_n(x)=\mathcal{H}_n(x)e^{x^2/2}$ with  
$\mathcal{H}_n(x)\equiv H_n(ix)$ 
correspond to non-physical (i.e. non-normalizable)
eigenstates of $H$ with eigenvalues $E^-_n=-(2n+1)$.
The corresponding functions
 $\widetilde{\psi_n}$ and  $\widetilde{\psi^-_n}$
are non-physical (non-normalizable) eigenfunctions of 
$H$ of eigenvalues $E_n=2n+1$ and 
$E_n=-(2n+1)$, $n=0,1,\dots$, respectively.
They will play  important role in the structure and properties 
of the REQHO system \cite{Adler,Dub,CPRS,FellSmi,Sesma,GGM,Pupas}. 

\begin{figure}[htbp]
\begin{center}
\includegraphics[scale=0.3]{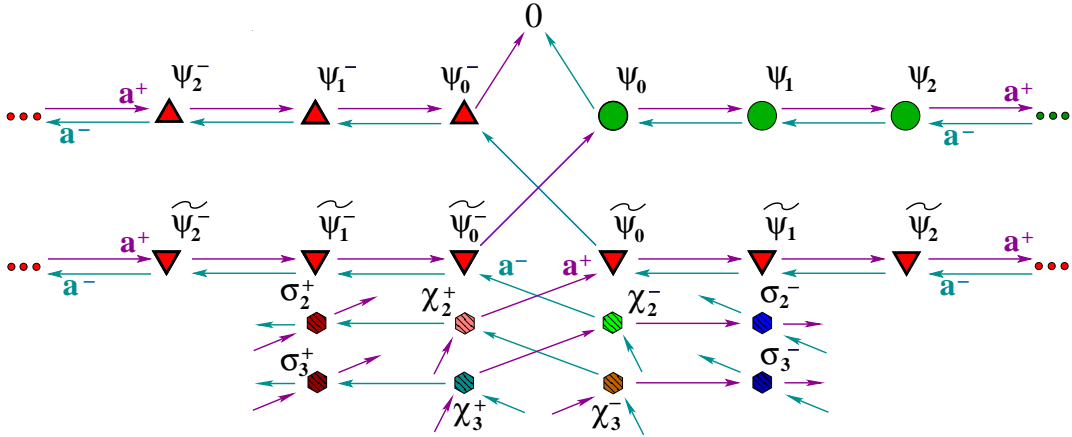}
\caption{Discrete flows of the ladder operators of the
QHO. Operator  $a^-$ acts left and up, and 
$a^+$ acts right and up. }
\label{Fig2}
\end{center}
\end{figure}

The well known peculiarity of the QHO system
in the context of the Darboux transformations  is that 
the choice of $E_*=1$, $\psi_*=\psi_0=e^{-x^2/2}$ 
gives $\mathcal{W}=-x$,  and  as the factorizing  operators $A_{\psi_*}$ and 
$A_{\psi_*}^\dagger$  we obtain  the usual, up to a multiplicative factor $\sqrt{2}$,
creation and annihilation operators,
\begin{equation}\label{a+a-}
a^-=
\frac{d}{dx}+x\,, 
\qquad
a^+=(a^-)^\dagger=-\frac{d}{dx}+x\,, \qquad
 [a^-,a^+]=2\,.
\end{equation}
In this case if $N=a^+\,a^-$  denotes the number operator for the QHO 
(with the spectrum $2n$, $n=0,1,\ldots$,  corresponding to 
a normalization
chosen in   (\ref{a+a-})) we have
 \begin{equation}
 [N,a^\pm]=\pm 2\, a^\pm\,,
 \qquad H=N+1\,.
 \label{hoNa}
 \end{equation}
 As a result, the  Darboux-partner system for the QHO 
 turns out to be 
 $\breve{H}=H+2$, which  is the same quantum harmonic oscillator 
but just with the spectrum of physical states shifted in $+2$. 
Since $\psi^-_0=1/\psi_0$, another choice 
$E_*=-1$, $\psi_*=\psi^-_0$ changes the role
of the Darboux-generating operators,
$A_{\psi_*}=a^+$, $A_{\psi_*}^\dagger=a^-$,
and  the partner system $\breve{H}=H-2$
in this second case is again 
the quantum harmonic oscillator 
but with the physical spectrum shifted in $-2$.
The action of the ladder operators 
on  physical and associated non-physical eigenstates 
of the QHO and on the related  Jordan and generalized 
Jordan states is described in Appendix. The corresponding 
discrete flows generated by $a^-$ and $a^+$ 
 are depicted in
Figure \ref{Fig2}.

In general, because of the  two-cyclic structure
associated with the Darboux transformations,
there appears a supersymmetry 
in the extended system  composed from $H$ and $\breve{H}$.
Since for the QHO with its equidistant spectrum 
the partner  generated   by the Darboux transformation 
based on the  eigenfunction $\psi_*=\psi_0$ 
(or  on  $\psi_*=\psi^-_0=1/\psi_0$) is
 the same system but just with the spectrum shifted
 exactly in one 
 energy step
 $\Delta E=+2$ (or, $\Delta E=-2$),
 the Darboux transformation generators 
 responsible for supersymmetric  structure 
 transmute into the ladder operators for 
the single harmonic oscillator system. 
Coherently with this, instead of 
a usual quantum mechanical supersymmetry
of the composed system, 
single quantum harmonic oscillator itself 
 is characterized by
the 
\emph{ bosonized} superconformal $\mathfrak{osp}(1\vert 2)$  
structure. 
The   $\mathfrak{osp}(1\vert 2)$   Lie superalgebra 
 is generated here by the  set of operators 
 \begin{equation}
 \mathcal{L}_\pm=\frac{1}{4}a^\pm\,,\qquad
J_0=\frac{1}{8}\{a^+,a^-\}=\frac{1}{4}H\,, \qquad
J_\pm=\frac{1}{4}(a^\pm)^2\,,
\label{ospGen}
\end{equation}
with nontrivial (anti)commutation
relations
\be\label{osp+}
\{\mathcal{L}_+,\mathcal{L}_-\}=\frac{1}{2}J_0\,,\qquad
\{\mathcal{L}_+,\mathcal{L}_+\}=\frac{1}{2}J_+\,,\qquad
\{\mathcal{L}_-,\mathcal{L}_-\}=\frac{1}{2}J_-\,,
\ee
\be\label{osp-}
[J_0,\mathcal{L}_\pm]=\pm\frac{1}{2}\mathcal{L}_\pm\,,\qquad
[J_+,\mathcal{L}_-]=-\mathcal{L}_+\,,\qquad
[J_-,\mathcal{L}_+]=\mathcal{L}_-\,,
\ee
\be\label{so(2,1)}
[J_0,J_\pm]=\pm J_\pm,,\qquad
[J_+,J_-]=-2J_0\,.
\ee
For this superalgebra a
reflection operator $\mathcal{R}=(-1)^{N/2}=e^{i\pi N/2}$ plays a role of a
$\Z_2$-grading operator, i.e.
\begin{equation}\mathcal{R}^2=1, \qquad \{\mathcal{R},\mathcal{L}_\pm\}=0,\qquad
[\mathcal{R},J_0]=[\mathcal{R},J_\pm]=0\,.\label{grading}
\end{equation}
 The operator 
\begin{equation}\label{ospCas}
\mathcal{C}_{\mathfrak{osp}(1\vert 2)}=-J_0^2+\frac{1}{2}(J_+J_-+J_-J_+)+2(\mathcal{L}_+\mathcal{L}_--
\mathcal{L}_-\mathcal{L}_+)
\end{equation}
is the quadratic Casimir  of
the  $\mathfrak{osp}(1\vert 2)$.

This corresponds to  the well known 
spectrum-generating superalgebra of the QHO  \cite{CromRit,defOSP},
on the  physical eigenstates $\psi_n(x)$ of which the 
infinite-dimensional irreducible representation of the $\mathfrak{osp}(1\vert 2)$
with $\mathcal{C}_{\mathfrak{osp}(1\vert 2)}=-\frac{1}{16}$ is realized.
The generators of the $\mathfrak{so}(2,1)$ Lie subalgebra (\ref{so(2,1)}) act  
irreducibly on the eigensubspaces of $\mathcal{R}$ 
spanned by the states $\psi_n(x)$ 
with even, $n=2n_+$, and odd, $n=2n_-+1$, $n_\pm=0,1,\dots$,
 values of $n$, where the operator
$J_0$ takes eigenvalues $n_++\frac{1}{4}$ and $n_-+\frac{3}{4}$, respectively.
On both these subspaces the $\mathfrak{so}(2,1)$ Casimir operator
$\mathcal{C}_{{}{\mathfrak{so}(2,1)}}=-J_0^2+\frac{1}{2}(J_+J_-+J_-J_+)$
takes the same value $\mathcal{C}_{{}{\mathfrak{so}(2,1)}}=\frac{3}{16}$.

In conclusion of this section we note 
 that a structure with a hidden bosonized  supersymmetry 
  \cite{bosSUSY,PVZ} 
also appears in periodic finite-gap and reflectionless quantum mechanical systems
\cite{BosHid}. There, however,  hidden supersymmetry
has a different origin   associated 
with  a presence of a nontrivial 
Lax-Novikov integral in the
 quantum mechanical systems
related to finite-gap and soliton solutions 
of the Korteweg-de Vries equation.

\section{Discrete flows in the REQHO and deformed 
superconformal $\mathfrak{osp}(1\vert 2)$ structure}

Take now a non-physical eigenstate 
$\psi^-_2(x)=(2x^2+1)e^{x^2/2}$ of the harmonic oscillator as the function
$\psi_*$ to generate  Darboux transformation.
This is a nodeless function, and the associated 
Darboux-transformed  system will be given by a non-singular on $\R$
potential.
For the sake of simplicity we denote by $A^-$  the corresponding first order operator 
(\ref{Adef}), in which the  superpotential
\begin{equation}
\mathcal{W}(x)=\frac{d}{dx}\left(\ln \psi^-_{2}\right)=x+\frac{4x}{2x^2+1}=
x+\frac{1}{x+\frac{i}{\sqrt{2}}}+\frac{1}{x-\frac{i}{\sqrt{2}}}\,
\end{equation} has simple poles at $\infty$ 
and $\pm \frac{i}{\sqrt{2}}$.
By  construction, $A^-\psi^-_{2}=0$,  and
 $A^+\left(\frac{1}{\psi^-_{{}_2}}\right)=0$,  where $A^+=(A^-)^\dagger$.
A simple computation gives 
\be\label{Hoscill}
A^+ A^-=-\frac{d^2}{dx^2}+x^2+5=N+6\equiv H_{{}_O},
\ee
\be\label{HdefO}
A^- A^+ =-\frac{d^2}{dx^2}+x^2+3+8\frac{2x^2-1}{(2x^2+1)^2}\equiv \breve{H}_{{}_{O}}\,,
\ee
where $N$ is the number operator for the QHO, 
$N=a^+a^-$.
Here $H_{{}_O}$ represents the QHO Hamiltonian shifted 
by an additive constant $5$. Hamiltonian operator 
$\breve{H}_{{}_{O}}$ describes the REQHO system 
with the physical bound states $\Psi_0=\frac{1}{\psi^-_2}=A^-\widetilde{\psi^-_2}$ 
and $\Psi_{n+1}=A^-\psi_n$  of energies $E_0=0$
and
$E_{n+1}=6+2n$, $n=0,1,\ldots$, 
constructed from  the corresponding QHO
states.

Let us  introduce  the third order differential operators
\be\label{dressedA}
\mathcal{A}^-=A^-a^-A^+\,,\qquad
\mathcal{A}^+=(\mathcal{A}^-)^\dagger=A^-a^+A^+\,.
\ee
These are the Darboux-dressed ladder operators of the QHO.
The operator $A^+$ maps a physical or non-physical eigenstate of $\breve{H}_{{}_O}$ 
into an eigenstate  (of the same nature) of the QHO,
to which $a^-$ or $a^+$ is then applied, and the obtained in this way eigenstate
of $H_{{}_O}$
is mapped  by $A^-$ into another
eigenstate 
of $\breve{H}_{{}_O}$.
Operators (\ref{dressedA})  satisfy the following commutation relations with the REQHO
Hamiltonian,
\be\label{ospdef0}
[\breve{H},\mathcal{A}^\pm]=\pm 2 \mathcal{A}^\pm\,,
\ee
for which from now on  we use a simplified notation
$\breve{H}$.
To find  (\ref{ospdef0}) we used the intertwining 
relations $A^+\breve{H}=H_{{}_O}A^+$,
$A^-H_{{}_O}=\breve{H} A^-$ as well as
Eq. (\ref{hoNa}).
Relation (\ref{ospdef0}) is generalized further for
\be\label{fHApm}
[\breve{H},\mathcal{A}^{\pm n}]=\pm 2n\mathcal{A}^{\pm n}\,,
\quad n=1,2,\ldots\,,
\ee
and 
$
f(\breve{H})\mathcal{A}^\pm=\mathcal{A}^\pm
f(\breve{H}\pm 2)
$
for an arbitrary polynomial function $f(\breve{H})$. 

The operators $\mathcal{A}^+$ and 
$\mathcal{A}^-$  transform 
eigenstates 
of $\breve{H}$, which are not from their kernels,
 into  
 eigenstates of $\breve{H}$ with the increased 
 and  decreased 
 in two energy values.
In this aspect they act analogously to
the  ladder operators $a^+$ and  $a^-$
in the QHO system.
There are, however,   essential differences.
These third order differential operators
satisfy relations
\be\label{A+A-H}
\mathcal{A}^+\mathcal{A}^-=\breve{H}(\breve{H}-2)(\breve{H}-6)\equiv 
\Phi(\breve{H})
\,,\qquad
\mathcal{A}^-\mathcal{A}^+=\breve{H}(\breve{H}+2)(\breve{H}-4)=
\Phi(\breve{H}+2)\,,
\ee
which 
 follow from (\ref{Hoscill}), (\ref{HdefO}),
(\ref{hoNa}) 
and intertwining properties of $A^\pm$,
and include 
the degree three polynomial $\Phi(\lambda)=\lambda (\lambda -2)(\lambda-6)$. 
{}From (\ref{A+A-H})   and (\ref{fHApm}) 
we also obtain the relations which will be used in what follows:
\be\label{AAadd1}
\mathcal{A}^{+2}\mathcal{A}^-=\Phi(\breve{H}-2)\mathcal{A}^+\,,\qquad
\mathcal{A}^-\mathcal{A}^{+2}=\Phi(\breve{H}+2)\mathcal{A}^+\,,
\ee
\be\label{AAadd2}
\mathcal{A}^{-2}\mathcal{A}^+=\Phi(\breve{H}+4)\mathcal{A}^-\,,\qquad
\mathcal{A}^+\mathcal{A}^{-2}=\Phi(\breve{H})\mathcal{A}^-\,,
\ee
\be\label{AAadd3}
\mathcal{A}^{+2}\mathcal{A}^{-2}=\Phi(\breve{H})\Phi(\breve{H}-2)\,,\qquad
\mathcal{A}^{-2}\mathcal{A}^{+2}=\Phi(\breve{H}+2)\Phi(\breve{H}+4)\,.
\ee
Both  third order polynomials $\Phi(\breve{H})$ and 
 $\Phi(\breve{H}+2)$  in (\ref{A+A-H})  include a factor  $\breve{H}$. 
  This reflects 
the essential peculiarity of the REQHO system: 
its ground-state $\Psi_0$  of zero energy 
is annihilated by both 
operators $\mathcal{A}^-$ and $\mathcal{A}^+$,
\be
\mathcal{A}^-\Psi_0=\mathcal{A}^+\Psi_0=0\,,
\ee
because  $\Psi_0=\frac{1}{\psi^-_2}$  is the kernel of $A^+$.

\begin{figure}[htbp]
\begin{center}
\includegraphics[scale=0.3]{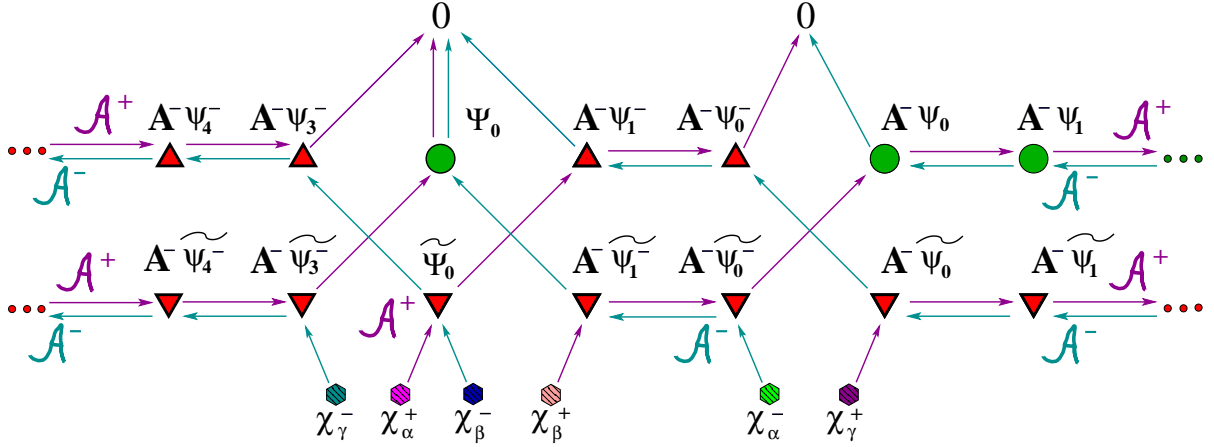}
\caption{Discrete flows of the ladder operators  of the REQHO.
Operator  $\mathcal{A}^-$ acts left and up, and 
$\mathcal{A}^+$ acts right and up. }
\label{Fig3}
\end{center}
\end{figure}

Consider  now other  properties 
of the lowering   ladder operator 
$\mathcal{A}^-$.
It also annihilates  the first excited physical state 
$\Psi_1(x)=A^-\psi_0(x)$, 
$
\mathcal{A}^-\Psi_1=0
$
due to sequential action of the operators 
$A^+$ and then $a^-$.
Moreover, it annihilates  a 
non-physical eigenstate $A^-\psi^-_1$ of $\breve{H}$
by means of  transforming it by the second order 
operator $a^-A^+$ into the kernel of $A^-$.
As the kernel of the
third-order differential operator 
$\mathcal{A}^-$ is three-dimensional, 
it is spanned by  the three eigenstates of $\breve{H}$,
\be\label{ker-1}
\ker\, \mathcal{A}^-={\rm span}\, \{\Psi_0, \,
A^-\psi^-_1,\,
\Psi_1\}\,,
\ee
whose eigenvalues $E=0, 2, 6$ correspond to zeros of the third degree 
polynomial $\Phi(\breve{H})$ in the first equality in (\ref{A+A-H}).
The operator $\mathcal{A}^-$ acts as a lowering ladder operator, and 
it is also interesting to look  for kernels of powers 
$(\mathcal{A}^-)^n$ with $n=2,3,\ldots$.  \emph{A priori} it is clear that due to 
the presence of another physical state  in the 
kernel of $\mathcal{A}^-$, which is   the first exited state $\Psi_1$
in the spectrum of $\breve{H}$,
and of the non-physical state 
$A^-\psi^-_1$ with eigenvalue $E=2$ located 
between the energies 
$E=0$ and $E=6$ of the physical zero modes 
of $\mathcal{A}^-$, 
some additional peculiarities 
have to appear in comparison with  the case of the QHO. 
Note first that (\ref{fHApm}) implies that  $\ker\,  (\mathcal{A}^-)^2$ must be invariant under
the action of $\breve{H}$.
On the other hand, we remark  
that $\ker\, \mathcal{A}^-\subset \ker\,(\mathcal{A}^-)^2$. 
Moreover, $\psi\in \ker\,(\mathcal{A}^-)^2$
if and only if $\mathcal{A}^-(\psi)\in \ker\,\mathcal{A}^-$, and 
therefore $\ker\,  (\mathcal{A}^-)^2$ is generated by 
 $\ker\,  \mathcal{A}^-$ and the pre-images under 
$ \mathcal{A}^-$ of $\Psi_0$, $A^-\psi^-_1$ and
$\Psi_1$, i.e. 
one  finds that
\be
{\rm ker}\, (\mathcal{A}^-)^2={\rm span}\, 
\{ {\rm ker}\, \mathcal{A}^-,\,  A^-\widetilde{\psi_1^-},\,  A^-\psi^-_0, \, \Psi_2\}\,.
\ee
Here $\Psi_2=A^-\psi_{1}$ is a physical eigenstate at the next 
energy level $E=8$, and two other states $A^-\widetilde{\psi_1^-}$ 
and $A^-\psi^-_0$ 
are non-physical eigenstates
of $\breve{H}$ of energies $E=2$ and $E=4$. Under  the action
of $\mathcal{A}^-$ the states $A^-\widetilde{\psi_1^-}$, 
$A^-\psi^-_0$  and $\Psi_2$ 
are transformed into the states  
$\Psi_0$, $A^-\psi^-_1$ and $\Psi_1$ from the kernel (\ref{ker-1}).
One can proceed  in this way and identify the  
action of the decreasing operator on all the physical eigenstates 
of the system and on the associated non-physical eigenstates
of the special form $\widetilde{\Psi_0}$,
$A^-\widetilde{\psi_n}$, $n=0,1,\ldots$,
and 
$A^-\psi^-_n$,
$A^-\widetilde{\psi^-_n}$, $n=0,1,3,4,5,\ldots$.
This action is depicted on Figure  \ref{Fig3}. 
The figure also shows that 
the pre-images of all the indicated physical and non-physical
eigenstates of $\breve{H}$ 
with eigenvalues $E_n=2n$, $n\in \Z$, 
are contained in the same set 
of eigenstates with the exception of the  
three non-physical 
states $A^-\widetilde{\psi^-_0}$, 
$\widetilde{\Psi_0}=\widetilde{\left(\frac{1}{\psi^-_2}\right)}$,
and $A^-\widetilde{\psi^-_3}$ of the 
eigenvalues $E=4$, $E=0$ and $E=-2$, respectively.
These  eigenvalues coincide with the set 
of zeros of the polynomial $\Phi(\breve{H}+2)$ 
that appears in the second relation in (\ref{A+A-H}).
The preimages of the indicated states are the 
states $\chi^-_a$, $a=\alpha,\beta,\gamma$, having the structure
\be\label{chia-}
\chi^-_a(x)=\frac{1}{\psi^-_2(x)}\int^x\psi_0(\xi)\psi^-_2(\xi)\left(\int^\xi \psi^-_0
(\eta)\rho^-_a(\eta)d\eta\right)d\xi\,.
\ee
Here 
$\rho^-_\alpha=\widetilde{\psi^-_0}$, 
$\rho^-_\beta=\Omega_{\psi^-_2}$, 
$\rho^-_\gamma=\widetilde{\psi^-_3}$,
and
$\mathcal{A}^-\chi^-_\alpha=A^-\widetilde{\psi^-_0}$,
$\mathcal{A}^-\chi^-_\beta=\widetilde{\Psi_0}$,
$\mathcal{A}^-\chi^-_\gamma=A^-\widetilde{\psi^-_3}$.
The states $\chi^-_a$ are not eigenstates of $\breve{H}$ but satisfy relations
\be
\breve{H}(\breve{H}-6)\chi^-_\alpha=A^-\psi_0\,,
\qquad
\breve{H}(\breve{H}-2)(\breve{H}-6)\chi^-_\beta=A^-\psi^-_1\,,
\qquad
\breve{H}(\breve{H}-6)\chi^-_\gamma=\Psi_0\,.
\ee
This implies  that the following polynomials in the Hamiltonian
$\breve{H}$ annihilate the states $\chi_a^-$:
\be\label{chi-Jordan}
\breve{H}(\breve{H}-6)^2\chi^-_\alpha=0\,,\qquad
\breve{H}(\breve{H}-2)^2(\breve{H}-6)\chi^-_\beta=0\,,\qquad
\breve{H}^2(\breve{H}-6)\chi^-_\gamma=0\,.
\ee
In correspondence with   (\ref{chi-Jordan}),
that  generalizes relations (\ref{Jordan}), we call
the states $\chi^-_a$ the  generalized Jordan states
of  the REQHO
since they are destroyed by the polynomials in $\breve{H}$ 
with different roots.
Note also that 
\be\label{A-4ker}
(\mathcal{A}^-)^3\chi^-_\alpha=\Psi_0 \quad
\Rightarrow\quad
\chi^-_\alpha\in{\ker}\,(\mathcal{A}^-)^4\,,
\ee
whereas $\chi^-_\beta$ and $\chi^-_\gamma$
are not annihilated by any degree of the ladder operator $\mathcal{A}^-$
and in this aspect they are similar
to Jordan state $\chi^-_2$ in the QHO
system, see Eq. (\ref{lam-}).

The kernel of the raising  ladder operator is
\be
{\ker}\, \mathcal{A}^+={\rm span}\, 
\{ A^-{\psi^-_3},\, \Psi_0,\,  A^-\psi^-_0\}\,.
\ee
The action  of 
$\mathcal{A}^+$ is illustrated  by the same 
 Figure \ref{Fig3}.
The corresponding shown there 
generalized Jordan states $\chi^+_a$,
$a=\alpha,\beta,\gamma$,  
are given by relations similar to (\ref{chia-}),
\be\label{chia+}
\chi^+_a(x)=\frac{1}{\psi^-_2(x)}\int^x\psi^-_0(\xi)\psi^-_2(\xi)\left(\int^\xi \psi_0
(\eta)\rho^+_a(\eta)d\eta\right)d\xi,
\ee
where 
$\rho^+_\alpha=\Omega_{\psi^-_2}$, 
$\rho^+_\beta=\widetilde{\psi^-_1}$, 
$\rho^+_\gamma=\widetilde{\psi_0}$,
and
$\mathcal{A}^+\chi^+_\alpha=\widetilde{\Psi_0}$,
$\mathcal{A}^+\chi^+_\beta=A^-\widetilde{\psi^-_1}$,
$\mathcal{A}^+\chi^+_\gamma=A^-\widetilde{\psi_0}$.
The non-physical eigenstates $\widetilde{\Psi_0}$, 
 $A^-\widetilde{\psi^-_1}$ 
 and $A^-\widetilde{\psi_0}$ 
of $\breve{H}$ appearing here
have eigenvalues $E=0$, $E=2$, $E=6$, respectively,
which correspond to zeros of the polynomial $\Phi(\breve{H})$ in (\ref{A+A-H}).
The states $\chi^+_a$ satisfy relations 
\be
\breve{H}(\breve{H}+2)(\breve{H}-4)\chi^+_\alpha=A^-\psi^-_3\,,
\qquad
\breve{H}(\breve{H}-4)\chi^+_\beta=\Psi_0\,,	\qquad
\breve{H}(\breve{H}-4)\chi^+_\gamma=A^-\psi^-_0\,.
\ee
As a consequence,
\be
\breve{H}(\breve{H}+2)^2(\breve{H}-4)\chi^+_\alpha=0\,,
\qquad
\breve{H}^2(\breve{H}-4)\chi^+_\beta=0\,,
\qquad
\breve{H}(\breve{H}-4)^2\chi^+_\gamma=0\,.
\qquad
\ee
Note that we have here
\be\label{A+4ker}
(\mathcal{A}^+)^3\chi^+_\alpha=A^-\psi^-_0 \quad
\Rightarrow\quad
\chi^+_\alpha\in{\rm ker}\,(\mathcal{A}^+)^4\,,
\ee
cf. (\ref{A-4ker}).
The notations for the generalized Jordan states $\chi^\pm_a$ 
are chosen so that the ordering in the lower index 
 $a=\alpha, \beta, \gamma$ 
in wave functions $\chi^+_a$ corresponds to the ordering 
in energies of the associated  non-physical eigenstates 
$\widetilde{\Psi_0}$,
$A^-\widetilde{\psi^-_1}$ and 
$A^-\widetilde{\psi_0}$.
Generalized Jordan state $\chi^-_\alpha$ 
is characterized by the property (\ref{A-4ker}) to be similar 
to the property (\ref{A+4ker}) for $\chi^+_\alpha$.
Under subsequent application of 
the ladder operator $\mathcal{A}^+$ to the state
$\chi^+_\beta$ and of the operator $\mathcal{A}^-$
to $\chi^-_\beta$, these states are lifted up 
to the highest horizontal level shown in Figure \ref{Fig3}
to which physical eigenstates do belong,
while the 
generalized Jordan states  $\chi^+_\gamma$
and $\chi^-_\gamma$ are lifted up by analogous 
action of the corresponding ladder operator 
to the lower
horizontal level where only non-physical
eigenstates of $\breve{H}$ do appear.
As in the case of the QHO system,
one can proceed and construct iteratively the net of the 
related generalized Jordan states by 
finding  the pre-images and images of 
the six basic generalized Jordan states $\chi^\pm_a$,
and of the generated in such a way new states 
under the sequential  action of the ladder operators
$\mathcal{A}^\pm$.

A complete isolation of the ground-state $\Psi_0$
 from other normalizable  eigenstates 
$\Psi_n$ with $n=1,2,\ldots$, reflects here the fact
that two irreducible representations of the 
polynomially deformed 
superconformal algebra $\mathfrak{osp}(1\vert 2)$ are realized 
on the physical bound states of  the REQHO
system.  
The operators 
$\breve{\mathcal{L}}_\pm=\frac{1}{4}\mathcal{A}^\pm$
can be identified as the  odd generators of the 
superalgebra, $\{\mathcal{R}, \breve{\mathcal{L}}_\pm\}=0$,
while $\breve{J}_0=\frac{1}{4}\breve{H}$ and 
$\breve{J}_\pm=\frac{1}{4}\mathcal{A}^{\pm 2}$
are its even generators,
$[\mathcal{R},\breve{J}_0]=[\mathcal{R},\breve{J}_\pm]=0$.
Here, as in the case of the QHO,
the operator
$\mathcal{R}=(-1)^{N/2}$ with $N=a^+ a^-$   is the $\Z_2$-grading operator,
$\mathcal{R}^2=1$.
The nontrivial commutation and anti-commutation relations of the deformed 
 $\mathfrak{osp}(1\vert 2)$ superalgebra of the REQHO 
can be found with the help of relations (\ref{ospdef0})--(\ref{AAadd3}). They can be  
presented in the form
\be\label{osp+!}
\{\breve{\mathcal{L}}_+,\breve{\mathcal{L}}_-\}=\frac{1}{2}
\mathcal{C}_{{}_{\mathcal{L}\mathcal{L}}}(\breve{J}_0)\, 
\breve{J}_0\,,\qquad
\{\breve{\mathcal{L}}_+,\breve{\mathcal{L}}_+\}=\frac{1}{2}\breve{J}_+\,,\qquad
\{\breve{\mathcal{L}}_-,\breve{\mathcal{L}}_-\}=\frac{1}{2}\breve{J}_-\,,\qquad
\ee
\be\label{osp-!!}
[\breve{J}_0,\breve{\mathcal{L}}_\pm]=\pm\frac{1}{2}\breve{\mathcal{L}}_\pm\,,\qquad
[\breve{J}_+,\breve{\mathcal{L}}_-]=-
\mathcal{C}_{{}_{J\mathcal{L}}}(\breve{J}_0)\, 
\breve{\mathcal{L}}_+\,,\qquad
[\breve{J}_-,\breve{\mathcal{L}}_+]=\breve{\mathcal{L}}_-\,\mathcal{C}_{{}_{J\mathcal{L}}}(\breve{J}_0)\,,
\ee
\be\label{so(2,1)!!}
[\breve{J}_0,\breve{J}_\pm]=\pm \breve{J}_\pm,,\qquad
[\breve{J}_+,\breve{J}_-]=-2 \mathcal{C}_{{}_{JJ}}(\breve{J}_0)\, 
\breve{J}_0\,.
\ee
The operator-valued  coefficients   
\begin{eqnarray}
&
\mathcal{C}_{{}_{\mathcal{L}\mathcal{L}}}(\breve{J}_0)=2\left(8 \breve{J}_0^2-10 \breve{J}_0 +1\right)\,,
\qquad
\mathcal{C}_{{}_{J\mathcal{L}}}(\breve{J}_0)=16
\left(\breve{J}_0-1\right)\left(3\breve{J}_0-1\right)\,,&\nonumber\\
&\mathcal{C}_{{}_{JJ}}(\breve{J}_0)=
16 \left(2\breve{J}_0-1\right)\left(\breve{J}_0-1\right)\left(24\breve{J}_0^2-14\breve{J}_0+7\right)
&\nonumber
\end{eqnarray}
appear here instead of the unit coefficients in the $\mathfrak{osp}(1\vert 2)$ superalgebra
(\ref{osp+}),
 (\ref{osp-}) and (\ref{so(2,1)})  of the QHO.
The ground-state $\Psi_0$ is annihilated by all the generators 
of the superalgebra and carries its  trivial one-dimensional 
representation.
On the higher bound states $\Psi_n$, $n=1,2,\ldots$, infinite-dimensional
 irreducible representation of the superalgebra is realized.
 The
 structure with two irreducible representations
 is reflected coherently in the discrete flows of the 
 ladder operators depicted  on  Figure \ref{Fig3}.
 
 In conclusion of this section we 
 note that the case of the deformed  $\mathfrak{osp}(1\vert 2)$ superalgebra of 
 the REQHO as well as the
 $\mathfrak{osp}(1\vert 2)$  Lie  superalgebra of the QHO system
 can be considered as particular cases of the algebra generated 
 by three elements $h$, $\alpha^+$ and $\alpha^-$
 subject to the relations 
 \be
 [h,\alpha^\pm]=\pm 2\alpha^\pm\,,\qquad
 \{\alpha^+,\alpha^-\}=F(h)+F(h+2)\,,
 \ee
 where $F(h)$ is some polynomial \cite{defOSP}.
 Such an algebra is characterized by the central element 
 \be
 \Xi=\alpha^{+2}\alpha^{-2}+\alpha^+\alpha^-\left(F(h)-F(h-2)\right)-(F(h))^2\,.
 \ee
 In the case of the QHO 
 we have a correspondence $\alpha^\pm=a^\pm$, $h=N+1$ and $F(h)=N$.
The  quadratic Casimir (\ref{ospCas})
 of the Lie superalgebra $\mathfrak{osp}(1\vert 2)$
  generated by the rescaled operators 
  $\alpha^\pm$, $\alpha^{\pm 2}$ 
 and $h$ 
 is nothing else as the rescaled and shifted for  additive constant 
 central element $\Xi$, $\mathcal{C}_{\mathfrak{osp}(1\vert 2)}=\frac{1}{16}(\Xi-1)$.
 For the REQHO system  operators  $\alpha^\pm$
 correspond to the ladder operators $\mathcal{A}^\pm$, 
 and we have $h=\breve{H}$, $F(h)=\Phi(\breve{H})$
 with $\Phi(\breve{H})$ defined in (\ref{A+A-H}). 
 The superalgebra (\ref{osp+!})--(\ref{so(2,1)!!}) 
 in this case can be considered as a polynomial deformation of 
 the $\mathfrak{osp}(1\vert 2)$ superalgebra.
 Using relations (\ref{A+A-H}) and (\ref{AAadd3}), one can easily check that 
 the central element $\Xi$ reduces here
  identically to zero.

\section{Conclusion and outlook}

To conclude, we list some problems to be interesting
for further investigation.

We have constructed  ladder operators for 
the simplest version of the 
REQHO system by the Darboux-dressing 
of creation and annihilation operators 
of the QHO.
This was done by means of the first order differential operators 
$A^-$ and $A^+$ 
which  intertwine the REQHO and QHO  Hamiltonians 
and factorize both of them. 
The applied  procedure here is  analogous  to the procedure
by which nontrivial Lax-Novikov integrals  
for  reflectionless quantum systems
are  constructed  by
the Darboux-dressing of  the free particle's momentum operator 
\cite{LaxNov}.
But the same REQHO system can  also be constructed
by means of the Darboux-Crum-Krein-Adler procedure
based on the usage of several eigenstates of the QHO.
In such a case  the intertwiners  
will be higher order differential operators.
One can expect that  the
existence of different 
Darboux and Darboux-Crum-Krein-Adler  transformations 
should reveal some new 
interesting aspects in 
the construction of the ladder operators 
for the  REQHO and related 
 dynamical symmetries 
 (spectrum generating algebras).

There exist   other rational extensions of the QHO system.
First, the analogs of the REQHO considered here can be 
generated by  taking non-physical nodeless eigenstate 
$\psi_{2n}^-$ with $n>1$ as a function $\psi_*$ 
to generate  Darboux transformations.
The ladder operators for such systems can be constructed in  a similar way,
by the Darboux-dressing of the ladder operators of the QHO.
We can generate also then a corresponding polynomially 
deformed bosonized $\mathfrak{osp}(1\vert 2)$ superalgebra,
whose trivial and infinite-dimensional representations will 
be realized on physical states of the corresponding 
rationally extended quantum harmonic oscillator.
It is interesting if there will be any essential difference 
in the structure of the discrete flows generated by the ladder operators 
in such systems
in comparison with the  REQHO system  considered here. 
The construction of the ladder operators
by taking into account the  existence of different 
Darboux-Crum-Krein-Adler transformations  to generate such systems
should also reveal a dependence 
on the order of the polynomial that presents
in the structure of the generating function
$\psi^-_{2n}(x)$ 
 and on a  size of the gap between 
the isolated ground-state  and the infinite tower of 
equidistant bound states.

A more complicated and a more rich picture  from the point of view of the ladder 
operators and related symmetries can be expected  
in rationally extended  quantum harmonic oscillator systems
with  number $l>1$ of isolated bound states in the spectrum. 
There, a priori two essentially  different cases should be distinguished.
One case is when 
$l>1$ bound states will be separated from  
the infinite tower of equidistant  bound states without any 
additional gaps
between those $l$ states. Another, more general case  is 
when isolated states include some additional
gaps between themselves.

It is known that Jordan states appear in confluent 
Darboux-Crum transformations \cite{Jord2}.
They, particularly,  were employed recently for the  design of
the PT-symmetric optical systems with invisible 
periodicity defects as well as completely invisible
 reflectionless PT-symmetric systems 
\cite{Jord1}.  It would be interesting to look 
for possible physical applications of 
the generalized Jordan states considered here. 

The considered 
REQHO system as well as its generalizations 
seem also to be  interesting from the point of view of possible 
physical applications since unlike other known 
deformations 
of the QHO, e.g.  related
to the  minimal length uncertainty relation \cite{Kempf,Rossi}, 
they provide a very specific change  of the spectrum. 
Namely, they  add effectively  a finite number of bound states 
in the lower part of the QHO spectrum, separated by an additional 
(\emph{adjustable}) gap,
without disturbing  the equidistant 
character of the rest of the infinite tower of the discrete levels.
In this aspect they are very similar, as it has been noted above
in another but related context,
to the quantum reflectionless
systems which add a finite number of discrete bound states 
into the spectrum of the free particle.
Such reflectionless systems are directly 
related to the soliton solutions to
the Korteweg-de Vries and modified Kortweg-de Vries 
equations, and   find a lot of 
interesting applications in very diverse areas of physics
including QCD, cosmology, 
solid states physics, the physics of polymers, plasma physics, and 
quantum optics, just to mention a few \footnote{See, e.g.,  \cite{ArPl,LaxNov,APtrans}
and references therein.}. 
Further results related to 
the ladder operators in 
rationally extended harmonic oscillator 
systems, which exploite the 
indicated similarity, will be presented elsewhere \cite{CarPlyprep}.

\vskip0.4cm

\noindent {\large{\bf Acknowledgements} } 
\vskip0.4cm

MSP thanks 
Ya. Ispolatov for discussions.
JFC and MSP acknowledge support from
research projects FONDECYT 1130017 (Chile),
Proyecto Basal USA1555 (Chile), 
MTM2015-64166-C2-1 (MINECO, Madrid) and DGA E24/1 (DGA, Zaragoza).
MSP is grateful  for the warm hospitality at  Zaragoza University.
JFC thanks for the kind hospitality at Universidad de Santiago de Chile.

\vskip1cm

\section{Appendix: Discrete chains of the states of the QHO}

We describe here  the action 
of the ladder operators  $a^+$ and $a^-$ 
on non-physical eigenstates $\widetilde{\psi_n}$
and $\widetilde{\psi_n^-}$
of the QHO and   the construction of
the associated Jordan and generalized Jordan states.

Making use of the identities $H'_n=2nH_{n-1}$
and $H_n=2xH_{n-1}-H'_{n-1}$  for Hermite polynomials, we obtain
the relations
\be\label{App1}
\int^x \frac{e^{\xi^2}}{H_n^2(\xi)} d\xi=-\frac{1}{2n}\int^x\frac{ e^{\xi^2}}{H_{n-1}(\xi)}d\left(\frac{1}{H_n(\xi)}\right)=
-\frac{e^{x^2}}{2nH_nH_{n-1}}+\frac{1}{2n}\int^x \frac{e^{\xi^2} }{H_{n-1}^2}d\xi\,,
\ee
from where  we find  that
\be\label{a-tilde}
a^-\widetilde{\psi_n}=\widetilde{\psi_{n-1}}\,,\qquad
n=1,2,\ldots.
\ee
Application  to both sides of this equality of the operator $a^+$
gives 
\be\label{a+tilde}
a^+\widetilde{\psi_n}=\widetilde{\psi_{n+1}}\,,\qquad
n=0,1,\ldots.
\ee
Changing  $x\rightarrow ix$ in (\ref{a-tilde}) and (\ref{a+tilde}), 
and taking into account that  $a^-\rightarrow ia^+$,
we also obtain
\be\label{a+tilde-}
a^+\widetilde{\psi^-_n}=\widetilde{\psi^-_{n-1}}\,,\quad
n=1,2,\ldots,\qquad
a^-\widetilde{\psi^-_n}=\widetilde{\psi^-_{n+1}}\,,\quad
n=0,1,\ldots.
\ee
In correspondence with (\ref{Apsi*})
and (\ref{A+psi*}),
\be
a^-\widetilde{\psi_0}=\frac{1}{\psi_0}=\psi^-_0\,,\qquad
a^+\widetilde{\psi^-_0}={\psi_0}\,.
\ee
We also have
\be\label{lam-}
\chi^-_2(x)=\psi_0(x)\int^x\psi^-_0(\xi)\widetilde{\psi^-_0}(\xi)d\xi\,,\qquad
a^-\chi^-_2(x)=\widetilde{\psi^-_0}\equiv \chi^+_1\,.
\ee
This is a Jordan state which obeys the relations
$
(H-1)\chi^-_2=\psi_0,
$
$
(H-1)^2\chi^-_2=0,
$
where $H=a^+a^-+1$. Analogously,
\be\label{lam+}
\chi^+_2(x)=\frac{1}{\psi_0}\int^x \psi_0(\xi)\widetilde{\psi_0}(\xi) d\xi\,,
\qquad
a^+\chi^+_2=\widetilde{\psi_0}\equiv \chi^-_1\,,
\ee
and 
$(H+1)\chi^+_2=\psi^-_0$,
$
(H+1)^2\chi^+_2=0.
$

Proceeding from the states $\chi^+_2$ and $\chi^-_2$,
one can construct an infinite net of 
related to them Jordan and generalized Jordan states.
First, as analogs of $\Omega_n$ and $\breve{\Omega}_n$ 
defined in (\ref{Omegan}) 
we have  the  states
$\chi^-_n$  and $\chi^+_n$, 
\be\label{chin-Def}
\chi^-_n(x)=\psi_0(x)\int^x\psi^-_0(\xi)\chi^+_{n-1}(\xi)d\xi\,,\qquad
\chi^+_n(x)=\psi_0^-(x)\int^x\psi_0(\xi)\chi^-_{n-1}(\xi)d\xi\,,
\ee
where the case $n=1$ is also  included by assuming
$\chi^-_0\equiv\psi_0$ and $\chi^+_0\equiv \psi^-_0$.
These are  the higher order Jordan states 
(\ref{Omegan}) 
generated on the basis of  $\psi_*=\psi_0$. They satisfy 
relations 
$a^-\chi^-_n=\chi^+_{n-1}$,
$a^+\chi^+_n=\chi^-_{n-1}$,  
and, consequently, 
$a^-(a^+a^-)^n\chi^-_{2n}=0$,  $a^+(a^-a^+)^n\chi^+_{2n}=0$.
Therefore, 
$(H-1)^{n+1}\chi^-_{k}=0$ and 
$(H+1)^{n+1}\chi^+_{k}=0$ for  $k=2n$, $2n+1$.

One can  define
 $\sigma^-_n=a^+\chi^-_n$, $\sigma^+_n=a^-\chi^+_n$,
 $n=2,\ldots$.
These are Jordan states obeying the relations 
 $(H-3)^{n}\sigma^-_{2n}=\psi_1$, 
$(H-3)^{n}\sigma^-_{2n+1}=\widetilde{\psi_1}$,
$(H+3)^n\sigma^+_{2n}=\psi^-_1$,
$(H+3)^n\sigma^+_{2n+1}=\widetilde{\psi^-_1}$, and, therefore,
$(H-3)^{n+1}\sigma^-_k=0$, 
$(H+3)^{n+1}\sigma^+_k=0$ for $k=2n$, $2n+1$. 

In the same vein  
the family of Jordan states 
 $\tau^-_n=a^+\sigma^-_n$ and  $\tau^+_n=a^-\sigma^+_n$,
 $n=2,\ldots$, 
 can be defined.
They satisfy the relations  $(H-5)^{n}\tau^-_{2n}=\psi_2$, 
$(H-5)^{n}\tau^-_{2n+1}=\widetilde{\psi_2}$,
$(H+5)^n\tau^+_{2n}=\psi^-_2$,
$(H+5)^n\tau^+_{2n+1}=\widetilde{\psi^-_2}$, 
and 
$(H-5)^{n+1}\tau^-_k=0$, 
$(H+5)^{n+1}\tau^+_k=0$ for $k=2n$, $2n+1$. 
These  discrete flows can be further continued `horizontally'.

 On the other hand, the states  defined via 
$\gamma^-_n=a^+\sigma^-_n$, $\gamma^+_n=a^-\sigma^+_n$,
$n=2,\ldots$, 
are reduced to  linear combinations of the already introduced 
Jordan states. Namely,  
$\gamma^-_{n}$ is  a linear combination of $\chi^-_n$ and $\chi^-_{n-2}$, 
and $\gamma^+_{n}$ is  a linear combination of
 $\chi^+_n$ and $\chi^+_{n-2}$.

Consider the states $\lambda^\pm_n$ given 
by means of  relations 
 $a^-\lambda^-_n=\chi^-_{n-1}$, 
$a^+\lambda^+_n=\chi^+_{n-1}$, $n=3,\ldots$.
The states $\lambda^-_n$  can be presented
in the form similar to that for $\chi^-_n$ 
in (\ref{chin-Def}) but with 
$\chi^+_{n-1}$ in the integrand changed for 
$\chi^-_{n-1}$. Analogously, 
$\lambda^+_n$  are  presented similarly to $\chi^+_n$ 
in   (\ref{chin-Def}) with 
$\chi^-_{n-1}$ in the integrand  changed for 
$\chi^+_{n-1}$.
For these states we have relations
 $(a^+a^-)^na^-\lambda^-_{2n}=0$, $(a^-a^+)^na^+\lambda^+_{2n}=0$,
$a^-(a^+a^-)^na^-\lambda^-_{2n+1}=0$, $a^+(a^-a^+)^na^+\lambda^+_{2n+1}=0$.
As a consequence, 
$(H-1)\lambda^-_n=\chi^+_{n-2}$, $(H+1)\lambda^+_n=\chi^-_{n-2}$.
Therefore, these are generalized Jordan states
which obey the relations
$(H-1)(H-3)^n\lambda^-_k=0$,
$(H+1)(H+3)^n\lambda^+_k=0$ with $k=2n$, $2n-1$.

Similarly,  generalized Jordan states $\mu^\pm_n(x)$
can be  defined  proceeding from the states  $\lambda^\pm_n$ via the relations 
$a^+\mu^-_n=\lambda^-_{n+1}$,
$a^-\mu^+_n=\lambda^+_{n+1}$, $n=2,\ldots$.
Then 
\be
\mu^-_n(x)=\psi^-_0(x)\int^x\psi_0(\xi)\lambda^-_{n+1}(\xi)d\xi\,,\qquad
\mu^+_n(x)=\psi_0(x)\int^x\psi^-_0(\xi)\lambda^+_{n+1}(\xi)d\xi\,.
\ee
For these states we have 
$\chi^-_n=(H+1)\mu^-_n$,
$\chi^+_n=(H-1)\mu^+_n$.
They are generalized Jordan states obeying the relations 
$(H+1)(H-1)^{n+1}\mu^-_k=0$, 
$(H-1)(H+1)^{n+1}\mu^+_k=0$
with $k=2n$, $2n+1$.

The described procedure of the construction of the 
Jordan and generalized Jordan states can be continued further in 
the obvious way.
The discrete flows corresponding to the action 
of the ladder operators on the physical and associated non-physical eigenstates
of the QHO Hamiltonian and associated Jordan 
and generalized Jordan states are illustrated by 
Figure \ref{Fig2}.



\begin{thebibliography}{99}


\bibitem{Darb}
G. Darboux,\emph{ ``Sur une proposition relative aux \'equations 
lin\'eaires,"}  C. R. Acad. Sci Paris {\bf 94} (1882) 1456.

\bibitem{Crum}
M. M. Crum, \emph{``Associated Sturm-Liouville systems,"}
\href{http://qjmath.oxfordjournals.org/content/6/1/121.full.pdf+html}{
Quart. J. Math. Oxford  {\bf 6} (1955) 121}.

\bibitem{Krein}
M. G. Krein,
\emph{ ``On a continuous analogue of a Christoffel 
 formula from the theory of orthogonal polynomials," }
 Dokl. Akad. Nauk SSSR {\bf 113} (1957) 970.
 
 \bibitem{Adler}
 V. E. Adler, \emph{``A modification of Crum's method,"}
    \href{http://link.springer.com/article/10.1007%2FBF01035458}{
 Theor. Math. Phys. {\bf 101} (1994) 1381}.


 \bibitem{Matveev}
V. B. Matveev and M. A. Salle, \textsl{Darboux Transformations and Solitons}
(Springer, Berlin,
1991).

\bibitem{Schr}
  E.~Schr\"odinger,
  \emph{``A method of determining quantum-mechanical eigenvalues and eigenfunctions,''}
  Proc.\ Roy.\ Irish Acad.\ (Sect.\ A) {\bf 46} (1940) 9.

\bibitem{InfHull} 
  L.~Infeld and T.~E.~Hull,
\emph{  ``The factorization method,''}
  \href{http://journals.aps.org/rmp/abstract/10.1103/RevModPhys.23.21}{
  Rev.\ Mod.\ Phys.\  {\bf 23} (1951)  21}.

\bibitem{CR00} J. F. Cari\~nena and  A. Ramos, 
 \emph{``Riccati equation, Factorization Method and Shape Invariance,"}
   \href{http://www.worldscientific.com/doi/abs/10.1142/S0129055X00000502}{
Rev.\  Math.\ Phys.\ {\bf 12} (2000)   1279}
\href{http://arxiv.org/abs/math-ph/9910020}{\textcolor{magenta}{[arXiv:math-ph/9910020]}}.

\bibitem{MelRos}
B. Mielnik and O. Rosas-Ortiz,
\emph{  ``Factorization: little or great algorithm?,''}
  \href{http://iopscience.iop.org/article/10.1088/0305-4470/37/43/001/meta}{
J. Phys. A  {\bf 37} (2004) 10007}. 

\bibitem{Witten1} 
  E.~Witten,
\emph{  ``Dynamical Breaking of Supersymmetry,''}
 \href{http://www.sciencedirect.com/science/article/pii/0550321381900067}{Nucl.\ 
 Phys.\ B {\bf 188} (1981)  513}.

\bibitem{Witten2}
  E.~Witten,
\emph{  ``Supersymmetry and Morse theory,''}
\href{http://projecteuclid.org/euclid.jdg/1214437492}{
  J.\ Diff.\ Geom.\  {\bf 17} (1982)  661}.

\bibitem{CoKhSu}
  F.~Cooper, A.~Khare and U.~Sukhatme,
\emph{   ``Supersymmetry and quantum mechanics,''}
\href{http://www.sciencedirect.com/science/article/pii/037015739400080M}{
  Phys.\ Rept.\  {\bf 251} (1995) 267}
  \href{http://arxiv.org/abs/hep-th/9405029}{\textcolor{magenta}{
  [hep-th/9405029]}}.
  

  \bibitem{CR01} J. F. Cari\~nena, D. J. Fern\'andez  and  A. Ramos, 
  \emph{  ``Group theoretical approach to the intertwined  Hamiltonians,''}
 \href{http://www.sciencedirect.com/science/article/pii/S0003491601961792}{
Ann.\  Phys.\  {\bf 292} (2001)  42}
\href{http://arxiv.org/abs/math-ph/0311029}{\textcolor{magenta}{[arXiv:math-ph/0311029]}}.

\bibitem{CR08} J. F. Cari\~nena and  A. Ramos, 
  \emph{  ``Generalized B\"acklund-Darboux transformations in one-dimensional 
quantum mechanics,''}
\href{http://www.worldscientific.com/doi/abs/10.1142/S0219887808002989}{
Int.\ J. Geom.\ Methods Mod. \ Phys.\ {\bf 05} (2008) 605}.

\bibitem{ArPl} 
  A.~Arancibia and M.~S.~Plyushchay,
\emph{    ``Chiral asymmetry in propagation of 
soliton defects in crystalline backgrounds,''}
\href{http://journals.aps.org/prd/abstract/10.1103/PhysRevD.92.105009}{
  Phys.\ Rev.\ D {\bf 92} (2015)  05009}
  \href{http://arxiv.org/abs/1507.07060}{\textcolor{magenta}{[arXiv:1507.07060 [hep-th]]}}.

\bibitem{VesSha}
A. P. Veselov  and A. B. Shabat,   
{\it ``Dressing chains and the spectral theory of the Schr\"odinger operator,"} 
\href{http://link.springer.com/article/10.1007%2FBF01085979}{
Funct. Anal. Appl. 27 (1993) 81}.  



\bibitem{NMPZ}  
 S. P. Novikov, S. V. Manakov, L. P. Pitaevskii, and V. E. Zakharov, 
  \textsl{Theory of Solitons} (Plenum, New York, 1984).

\bibitem{GesHol}
 F. Gesztesy and H. Holden,
   \textsl{Soliton Equations and their Algebro-Geometric Solutions}, (Cambridge Univ. Press, 2003). 
  
\bibitem{AdlPai}
 V. E. Adler,  {\it ``Nonlinear chains and Painlev\'e equations,"}
 \href{http://www.sciencedirect.com/science/article/pii/016727899490104X}{
 Physica D 73 (1994) 335}.
  
\bibitem{Hiet}
R. Willox and J. Hietarinta, {\it ``Painlev\'e equations from Darboux chains: I.  PIII - PV,"}
 \href{http://iopscience.iop.org/article/10.1088/0305-4470/36/42/014}{
J. Phys. A {\bf 36} (2003) 10615}.  

\bibitem{Iso1}
J. J. Duistermaat and F. A. Gr\"unbaum, {\it ``Differential equations in the spectral parameter,"}
\href{http://link.springer.com/article/10.1007/BF01206937}{
Comm. Math. Phys. {\bf 103}  (1986) 177}.

\bibitem{Iso2}
A. Oblomkov, {\it ``Monodromy-free Schr\"odinger operators with quadratically increasing potentials,"}
\href{http://link.springer.com/article/10.1007%2FBF02557204}{
Theor. Math. Phys. {\bf 121}  (1999) 1574}.

\bibitem{Dub}
S. Yu. Dubov, V. M. Eleonskii and N. E. Kulagin, 
 \emph{  ``Equidistant spectra of anharmonic oscillators,"}  
 \href{http://www.jetp.ac.ru/cgi-bin/e/index/r/102/3/p814?a=list}{
Zh. Eksp. Teor. Fiz. 102 (1992)  814  [Sov. Phys. JETP 75, 446 (1992)]};
\href{http://scitation.aip.org/content/aip/journal/chaos/4/1/10.1063/1.166056}{
Chaos, {\bf 4} (1994) 47}.

\bibitem{CPRS} 
J. F. Cari\~nena, A. M. Perelomov, M. F. Ra\~nada and M. Santander, 
 \emph{``A quantum exactly solvable nonlinear oscillator related
to the isotonic oscillator,"}
\href{http://iopscience.iop.org/article/10.1088/1751-8113/41/8/085301/meta}{
 J. Phys. A: Math. Theor. {\bf 41} (2008) 085301}
 \href{http://arxiv.org/abs/0711.4899}{\textcolor{magenta}{  [arXiv:0711.4899 [quant-ph]]}}.


\bibitem{FellSmi}
J. M. Fellows and R. A. Smith, 
 \emph{  ``Factorization solution of a family of quantum nonlinear oscillators,"}
 \href{http://iopscience.iop.org/article/10.1088/1751-8113/42/33/335303/meta}{
  J. Phys. A {\bf 42} (2009) 335303}.
  
  \bibitem{Exc1}
  D. G\'omez-Ullate, N. Kamran, and  R. Milson,
 \emph{  ``An extended class of orthogonal polynomials defined by a Sturm-Liouville problem,"}
\href{http://www.sciencedirect.com/science/article/pii/S0022247X09004569}{
 J. Math. Anal. Appl. {\bf 359}
(2009) 352}
\href{https://arxiv.org/abs/0807.3939}{\textcolor{magenta}{[arXiv:0807.3939 [math-ph]]}};
 \emph{    ``An extension of Bochner's problem: exceptional invariant subspaces,"} 
\href{http://www.sciencedirect.com/science/article/pii/S0021904509001853}{
J. Approx. Theory {\bf 162} (2010) 987}
\href{https://arxiv.org/abs/0805.3376}{\textcolor{magenta}{[arXiv:0805.3376 [math-ph]]}}.
  
\bibitem{OdSas}
 S. Odake and  R. Sasaki, 
 \emph{    ``Infinitely many shape invariant potentials and new 
 orthogonal polynomials, }
 \href{http://www.sciencedirect.com/science/article/pii/S0370269309009186}{
 Phys. Lett. B {\bf 679} (2009) 414}
 \href{https://arxiv.org/abs/0906.0142}{\textcolor{magenta}{[arXiv:0906.0142 [math-ph]]}};
 \emph{    ``Another set of infinitely many exceptional (X) 
 Laguerre polynomials, }
 \href{http://www.sciencedirect.com/science/article/pii/S0370269310000158}{
Phys. Lett. B {\bf 684} (2010) 173}
\href{https://arxiv.org/abs/0911.3442}{\textcolor{magenta}{[arXiv:0911.3442 [math-ph]]}}.

\bibitem{SasTsZh}
R. Sasaki, S. Tsujimoto and  A. Zhedanov, 
 \emph{    ``Exceptional Laguerre and Jacobi polynomials 
and the corresponding potentials through Darboux-Crum
 transformations,"}
 \href{http://iopscience.iop.org/article/10.1088/1751-8113/43/31/315204/meta}{
J. Phys. A {\bf 43} (2010) 315204}
\href{https://arxiv.org/abs/1004.4711}{\textcolor{magenta}{[arXiv:1004.4711 [math-ph]]}}.


\bibitem{Ques}
C. Quesne, 
 \emph{   ``Exceptional orthogonal polynomials, exactly 
 solvable potentials and supersymmetry,"} 
 \href{http://iopscience.iop.org/article/10.1088/1751-8113/41/39/392001/meta}{
J. Phys. A {\bf 41} (2008) 392001} 
\href{http://arxiv.org/abs/0807.4087}{\textcolor{magenta}{[arXiv:0807.4087 [quant-ph]]}};
 \emph{   ``Solvable rational potentials and exceptional 
orthogonal polynomials in supersymmetric quantum mechanics,"} 
\href{http://www.emis.de/journals/SIGMA/2009/084/}{
SIGMA {\bf 5}  (2009) 084}
\href{https://arxiv.org/abs/0906.2331}{\textcolor{magenta}{[arXiv:0906.2331 [math-ph]]}}.

\bibitem{Grand}
Y. Grandati, 
 \emph{   ``Solvable rational extensions of the isotonic oscillator,"}
 \href{http://www.sciencedirect.com/science/article/pii/S000349161100039X}{
Ann. Phys. {\bf  326} (2011)  2074}
\href{https://arxiv.org/abs/1101.0055}{\textcolor{magenta}{[arXiv:1101.0055 [math-ph]]}}.

\bibitem{Sesma}
J. Sesma, 
 \emph{  ``The generalized quantum isotonic oscillator,"} 
 \href{http://iopscience.iop.org/article/10.1088/1751-8113/43/18/185303/meta}{
J. Phys.  A
{\bf 43} (2010) 185303} 
 \href{http://arxiv.org/abs/1005.1227}{\textcolor{magenta}{[arXiv:1005.1227 [quant-ph]]}}.

 
\bibitem{GGM}
D. G\'omez-Ullate, Y. Grandati, and R. Milson, 
 \emph{  ``Rational extensions of
the quantum harmonic oscillator and exceptional Hermite polynomials,"}
\href{http://iopscience.iop.org/article/10.1088/1751-8113/47/1/015203/meta}{
J. Phys. A {\bf 47}  (2014) 015203}
 \href{http://arxiv.org/abs/1306.5143}{\textcolor{magenta}{[arXiv:1306.5143 [maph-ph]]}}.
 
 \bibitem{Pupas}
 A. M. Pupasov-Maksimov,
 \emph{``Propagators of isochronous an-harmonic oscillators 
 and Mehler formula for the exceptional Hermite polynomials,"}
\href{http://www.sciencedirect.com.ezproxy.usach.cl/science/article/pii/S0003491615003565}{
Annals of Physics {\bf 363} (2015) 122}
\href{https://arxiv.org/abs/1502.01778}{\textcolor{magenta}{[arXiv:1502.01778 [maph-ph]]}}.

\bibitem{AsoCar}
  M.~Asorey, J.~F.~Carinena, G.~Marmo and A.~Perelomov,
 \emph{  ``Isoperiodic classical systems and their quantum counterparts,''}
\href{http://www.sciencedirect.com/science/article/pii/S0003491606001461}{
Annals Phys.\  {\bf 322} (2007) 1444}
\href{https://arxiv.org/abs/0707.4465}{\textcolor{magenta}{  [arXiv:0707.4465 [hep-th]]}}.

\bibitem{CPR07} J. F. Cari\~nena, A. M. Perelomov  and  M. F. Ra\~nada,
 \emph{ ``Isochronous classical systems and quantum systems with equally 
spaced spectra,''}, 
{\rm  In:}  {\sl Particles and Fields: Classical and Quantum},
\href{http://iopscience.iop.org/article/10.1088/1742-6596/87/1/012007}{Journal of Physics: Conference Series {\bf 87} (2007) 012007}.

\bibitem{CromRit}
M. de Crombrugghe and V. Rittenberg,
\emph{  ``Supersymmetric quantum mechanics,}
 \href{http://www.sciencedirect.com/science/article/pii/0003491683903160}
{Annals of Physics
{\bf 151} (1983) 99}.

\bibitem{defOSP} 
  J.~Van der Jeugt and R.~Jagannathan,
\emph{ ``Polynomial deformations of osp(1/2) and generalized parabosons,''}
 \href{http://scitation.aip.org/content/aip/journal/jmp/36/8/10.1063/1.530904}{ J.\ Math.\ Phys.\  {\bf 36} (1995)  4507}
  \href{http://arxiv.org/abs/hep-th/9410145}{\textcolor{magenta}{ [hep-th/9410145]}}.


\bibitem{KriLech}
S. Krivonos and O. Lechtenfeld,
\emph{``SU(2) reduction in $\mathcal{N}=4$ supersymmetric mechanics,"}
 \href{http://journals.aps.org/prd/abstract/10.1103/PhysRevD.80.045019}{
Phys. Rev. D {\bf 80} (2009) 045019}
\href{https://arxiv.org/abs/0906.2469}{\textcolor{magenta}{[arXiv:0906.2469 [hep-th]]}}.
 
 
\bibitem{FedLuk}
S. Fedoruk and J. Lukierski,
\emph{ ``Algebraic structure of Galilean superconformal symmetries,"}
 \href{http://journals.aps.org/prd/abstract/10.1103/PhysRevD.84.065002}{Phys. Rev. D {\bf 84}
 (2011) 065002}
\href{http://arxiv.org/abs/1105.3444}{\textcolor{magenta}{[arXiv:1105.3444 [math-ph]]}}.


\bibitem{PlyWi}
  M.~S.~Plyushchay and A.~Wipf,
  \emph{  ``Particle in a self-dual dyon background: hidden free nature, and exotic superconformal symmetry,''}
 \href{http://journals.aps.org/prd/abstract/10.1103/PhysRevD.89.045017}{ Phys.\ Rev.\ D {\bf 89} (2014) 045017} 
 \href{http://arxiv.org/abs/1311.2195}{\textcolor{magenta}{[arXiv:1311.2195 [hep-th]]}}.

 \bibitem{TDB} 
G. F. de T\'eramond, H. G. Dosch and S. J. Brodsky,
\emph{  ``Baryon spectrum from superconformal quantum 
mechanics and its light-front holographic embedding,"}
 \href{http://journals.aps.org/prd/abstract/10.1103/PhysRevD.91.045040}{Phys. Rev. D 
 {\bf 91} (2015) 045040}
 \href{http://arxiv.org/abs/1411.5243}{\textcolor{magenta}{ [arXiv:1411.5243 [hep-ph]]}}.

 \bibitem{IST} 
E. Ivanov, S. Sidorov and F. Toppan,
\emph{``Superconformal mechanics in $SU(2|1)$ superspace,"}
 \href{http://journals.aps.org/prd/abstract/10.1103/PhysRevD.91.085032}{Phys. Rev. D 
 {\bf 91} (2015) 085032 }
 \href{https://arxiv.org/abs/1501.05622}{\textcolor{magenta}{[arXiv:1501.05622 [hep-th]]}}.


\bibitem{Jord1} 
  F.~Correa, V.~Jakubsky and M.~S.~Plyushchay,
\emph{   ``$PT$-symmetric invisible defects and confluent 
Darboux-Crum transformations,''}
\href{http://journals.aps.org/pra/abstract/10.1103/PhysRevA.92.023839}{
  Phys.\ Rev.\ A {\bf 92} (2015) 023839}
 \href{http://arxiv.org/abs/1506.00991}{\textcolor{magenta}{  [arXiv:1506.00991 [hep-th]]}}.

\bibitem{Jord2} 
  A.~Schulze-Halberg,
\emph{  ``Wronskian representation for confluent supersymmetric transformation 
chains of arbitrary order,''}
  \href{http://link.springer.com/article/10.1140/epjp/i2013-13068-2}{
  Eur.\ Phys.\ J.\ Plus {\bf 128} (2013) 68}.

\bibitem{bosSUSY}
M.~S.~Plyushchay,
\emph{``Deformed Heisenberg algebra, fractional spin fields and 
supersymmetry without fermions,''}
\href{http://www.sciencedirect.com/science/article/pii/S0003491696900123}{
 Annals Phys.\  {\bf 245} (1996)  339}
 \href{http://arxiv.org/abs/hep-th/9601116}{\textcolor{magenta}{  [hep-th/9601116]}};
 \emph{ ``Hidden nonlinear supersymmetries in pure parabosonic systems,''}
 \href{http://www.worldscientific.com/doi/abs/10.1142/S0217751X00001981}{
  Int.\ J.\ Mod.\ Phys.\ A {\bf 15} (2000)  3679}
  \href{http://arxiv.org/abs/hep-th/9903130}{\textcolor{magenta}{ [hep-th/9903130]}};
  V.~Jakubsky, L.~M.~Nieto and M.~S.~Plyushchay,
\emph{  ``The origin of the hidden supersymmetry,''}
\href{http://www.sciencedirect.com/science/article/pii/S0370269310008270}{
  Phys.\ Lett.\ B {\bf 692} (2010) 51}
 \href{http://arxiv.org/abs/1004.5489}{\textcolor{magenta}{  [arXiv:1004.5489 [hep-th]]}}.

\bibitem{PVZ} 
  S.~Post, L.~Vinet and A.~Zhedanov,
 \emph{  ``Supersymmetric Quantum Mechanics with Reflections,''}
\href{http://iopscience.iop.org/article/10.1088/1751-8113/44/43/435301/meta}{J.\ Phys.\ A {\bf 44} (2011)  435301}
 \href{http://arxiv.org/abs/1107.5844}{\textcolor{magenta}{[arXiv:1107.5844 [math-ph]]}};
V. X. Genest, L. Vinet, Guo-Fu Yu, and  A. Zhedanov,
 \emph{``Supersymmetry of the quantum rotor,"}
 \href{http://arxiv.org/abs/1607.06967}{\textcolor{magenta}{[arXiv:1607.06967 [math-ph]]}}.

\bibitem{BosHid} 
  F.~Correa and M.~S.~Plyushchay,
\emph{   ``Hidden supersymmetry in quantum bosonic systems,''}
  \href{http://www.sciencedirect.com/science/article/pii/S0003491606002831}{
  Annals Phys.\  {\bf 322} (2007) 2493}
   \href{https://arxiv.org/abs/hep-th/0605104}{\textcolor{magenta}{[hep-th/0605104]}};
 F.~Correa, L.~M.~Nieto and M.~S.~Plyushchay,
 \emph{ ``Hidden nonlinear supersymmetry of finite-gap Lam\'e equation,''}
   \href{http://www.sciencedirect.com/science/article/pii/S0370269306014274}{
  Phys.\ Lett.\ B {\bf 644} (2007) 94}
 \href{https://arxiv.org/abs/hep-th/0608096}{\textcolor{magenta}{[hep-th/0608096]}};
 F.~Correa, V.~Jakubsky, L.~M.~Nieto and M.~S.~Plyushchay,
  \emph{ ``Self-isospectrality, special supersymmetry, and their effect 
  on the band structure,''}
  \href{http://journals.aps.org/prl/abstract/10.1103/PhysRevLett.101.030403}{Phys.\ Rev.\ Lett.\  {\bf 101} (2008) 030403}
  \href{http://arxiv.org/abs/0801.1671}{\textcolor{magenta}{[arXiv:0801.1671 [hep-th]]}}.
  

\bibitem{LaxNov} 
  A.~Arancibia, J.~Mateos Guilarte and M.~S.~Plyushchay,
 \emph{  ``Effect of scalings and translations on the supersymmetric 
 quantum mechanical structure of soliton systems,''}
  \href{http://journals.aps.org/prd/abstract/10.1103/PhysRevD.87.045009}{
  Phys.\ Rev.\ D {\bf 87} (2013) 045009}
 \href{http://arxiv.org/abs/1210.3666}{\textcolor{magenta}{  [arXiv:1210.3666 [math-ph]]}}.

\bibitem{Kempf}
 A.~Kempf,
  \emph{ ``Uncertainty relation in quantum mechanics with quantum group symmetry,''}
  \href{http://scitation.aip.org/content/aip/journal/jmp/35/9/10.1063/1.530798}{J.\ Math.\ Phys.\  {\bf 35} (1994) 4483}
 \href{http://arxiv.org/abs/hep-th/9311147}{\textcolor{magenta}{  [hep-th/9311147]}};
A. Kempf, G. Mangano and R. B. Mann,
 \emph{ ``Hilbert space representation of the minimal length uncertainty relation,"}
 \href{http://journals.aps.org/prd/abstract/10.1103/PhysRevD.52.1108}{
 Phys. Rev. D {\bf 52} (1995) 1108}
 \href{http://arxiv.org/abs/hep-th/9412167}{\textcolor{magenta}{  [hep-th/9412167]}}.

\bibitem{Rossi} 
  M.~A.~C.~Rossi, T.~Giani and M.~G.~A.~Paris,
  \emph{ ``Probing deformed quantum commutators,''}
\href{http://journals.aps.org/prd/abstract/10.1103/PhysRevD.94.024014}{Phys.\ Rev.\ D {\bf 94} (2016) 024014}
 \href{https://arxiv.org/abs/1606.03420}{\textcolor{magenta}{[arXiv:1606.03420 [quant-ph]]}}.

\bibitem{APtrans} 
  A.~Arancibia and M.~S.~Plyushchay,
  \emph{  ``Transmutations of supersymmetry through soliton scattering
   and self-consistent condensates,''}
\href{http://journals.aps.org/prd/abstract/10.1103/PhysRevD.90.025008}{Phys.\ Rev.\ D {\bf 90} (2014) 025008}
  \href{https://arxiv.org/abs/1401.6709}{\textcolor{magenta}{[arXiv:1401.6709 [hep-th]]}}.

\bibitem{CarPlyprep}
J. F. Cari\~nena and M. S. Plyushchay,
in preparation.

\end{thebibliography}
\end{document}